\documentclass{PoS}

\def\aj{AJ}%
\def\apj{ApJ}%
\def\apjs{ApJS}%
\def\mnras{MNRAS}%
\def\hi{H\,{\sc i}}
\def\m20{$M_{20}$}

\usepackage{url}
\usepackage{natbib}

\title{Quantified Morphology of HI Disks in the Universe}

\ShortTitle{Quantified HI Disk Morphology }

\author{\speaker{B. W. Holwerda}, W. J. G. de Blok, A. Bouchard, S-L. Blyth, K. van der Heyden, \\
        Department of Astronomy, University of Cape Town, Private Bag X3, 7700 Rondebosch, Republic of South Africa\\
        E-mail: \email{holwerda@ast.uct.ac.za}}
\author{N. Prizkal\\
Space Telescope Science Institute, Baltimore, MD 21218, USA\\}

\abstract{
The upcoming new perspective of the high redshift Universe in the 21 cm line of atomic hydrogen opens possibilities to explore topics of spiral disk evolution, hitherto reserved for the optical regime. 
The growth of spiral gas disks over Cosmic time can be explored with the new generation of radio telescopes, notably the SKA, and its precursors, as accurately as with the Hubble Space Telescope for stellar disks. Since the atomic hydrogen gas is the building block of these disks, it should trace their formation accurately. 

Morphology of \hi \ disks can now equally be quantified over Cosmic time. In studies of HST deep fields, the optical or UV morphology of high-redshift galaxy disks have been characterised using a few quantities: concentration (C), asymmetry (A), smoothness (S), second-order-moment ($M_{20}$), the GINI coefficient (G), and Ellipticity (E). We have applied these parameters across wavelengths and compared them to the HI morphology over the THINGS sample. NGC 3184, an unperturbed disk, and NGC 5194, the canonical 3:1 interaction, serve as examples for quantified morphology.

We find that morphology parameters determined in \hi \ are as good or better a tracer of interaction compared to those in any other wavelength, notably in Asymmetry, Gini and $M_{20}$.  This opens the possibility of using them in the parameterization pipeline for SKA precursor catalogues to select interacting or harassed galaxies from their \hi \ morphology. Asymmetry, Gini and $M_{20}$ may be redefined for use on data-cubes rather than \hi \ column density image. 

}

\FullConference{Panoramic Radio Astronomy: Wide-field 1-2 GHz research on galaxy evolution - PRA2009\\
		 June 02 - 05 2009\\
		 Groningen, the Netherlands}

\begin{document}

\section{Introduction}

In the coming years, the SKA precursors, MeerKAT, ASKAP, WSRT/APERTIF and EVLA will take 21 cm line emission data of a large number of nearby and distant galaxies. The resulting data-cubes and zero-order maps  will need automated schemes for source detection and characterisation. The morphology of \hi \ disks is greatly informed by recent interaction \citep{Hibbard01} and harassment by passing dark matter haloes \citep[][]{Kazantzidis08}. 
Therefore an automated morphology classification of \hi \ data is desirable, to comb the future catalogues of \hi \ surveys for interacting or perturbed galaxies.
We propose to adapt morphological parameters currently in use to characterise UV/optical data of nearby and distant galaxies for use on \hi \ data in order identify different types of galaxy disks.

\section{Quantified Morphology}

The morphology of spiral galaxies has been quantified in Hubble Space Telescope images of distant galaxies and reference nearby galaxy samples in the restframe UV and optical using two parameter schemes or a combination thereof. After the initial work by \cite{Abraham94, Abraham96a}, the Concentration-Asymmetry-Smoothness (CAS) space was established by \cite{CAS} and the Gini and Asymmetry parameters by \cite{Lotz04}. \cite{Scarlata07} added ellipticity for Hubble type classification. 
The benefits of quantified morphology parameters are that there is no need for an observer, they can be applied to rest-frame UV images, which trace the triggered star-formation in interacting galaxies and the galaxies have an intrinsically high surface brightness, aiding their detection. 

The parameters used are:\\

\noindent \underline{Concentration:} $C ~ = ~ 5 ~ log({r_{80} \over r_{20}})$, \\
where $r_f$ is the radius of the circular aperture containing $f$ percent of the light.\\

\noindent \underline{Asymmetry:} $A = {(I - I_{180}) \over I}$,\\
 where $I$ is the image, $I_{180}$ is the image rotated by $180^{\circ}$ around the galaxy's center\\

\noindent \underline{Smoothness:} $S = {(I - I_{S}) \over I}$, \\
where $I$ is the image, $I_{S}$ is the image smoothed with a certain kernel. The 0.2 Petrosian radius boxcar is in use, but we used a 6" Gaussian. This is effectively a parameterized version of unsharp masking \citep{Takamiya99,Malin97b}.\\

\noindent \underline{Gini:} $G = {1\over \bar{I} n (n-1)} \Sigma_i (2i - n - 1) I_i $, \\
where $I_i$ is the value of pixel i in an ordered list, $n$ is the number of pixels in the galaxy image, and $\bar{I}$ is the mean pixel value in the image. The Gini parameter is an index of equality in economics, with G=1 is perfectly unequal (all the flux in one pixel) and G=0 is perfectly equal (all pixel values the same).\\

\noindent \underline{$M_{20}$:} the relative contribution of the 20\% brightest pixels to the second order moment of the light in the image. The second order moment of pixel i is defined as: $M_i = I_i [(x_i - x_c)^2 + (y_i - y_c)^2]$, where $I_i$ is the value of pixel i in the image, and $x_i$ and $y_i$ are the x and y coordinates of that pixel and $x_c$ and $y_c$ are the position of the galaxy's centre.
The total second order is: $M_{tot} = \Sigma I_i [(x_i - x_c)^2 + (y_i - y_c)^2]$, and $M_{20}$ is defined as:
$M_{20} = log \left( {\Sigma_i^m M_i  \over  M_{tot}}\right)$, where $m$ is the 20\% brightest pixels.\\

\noindent \underline{Ellipticity}: $E = 1 - B/A$ where A and B are the major and minor axes of the image.\\

We note that many of these parameters need a predefined central position of the galaxy and all need a definition of which pixels in an image belong to the object (a segmentation map). We use the areas on the sky defined by \hi \ contours to determine which pixels belong to our target galaxy and the dynamical centres reported in \cite{Walter08}. 

\begin{figure*}
\centering
\includegraphics[angle=0, width=\textwidth]{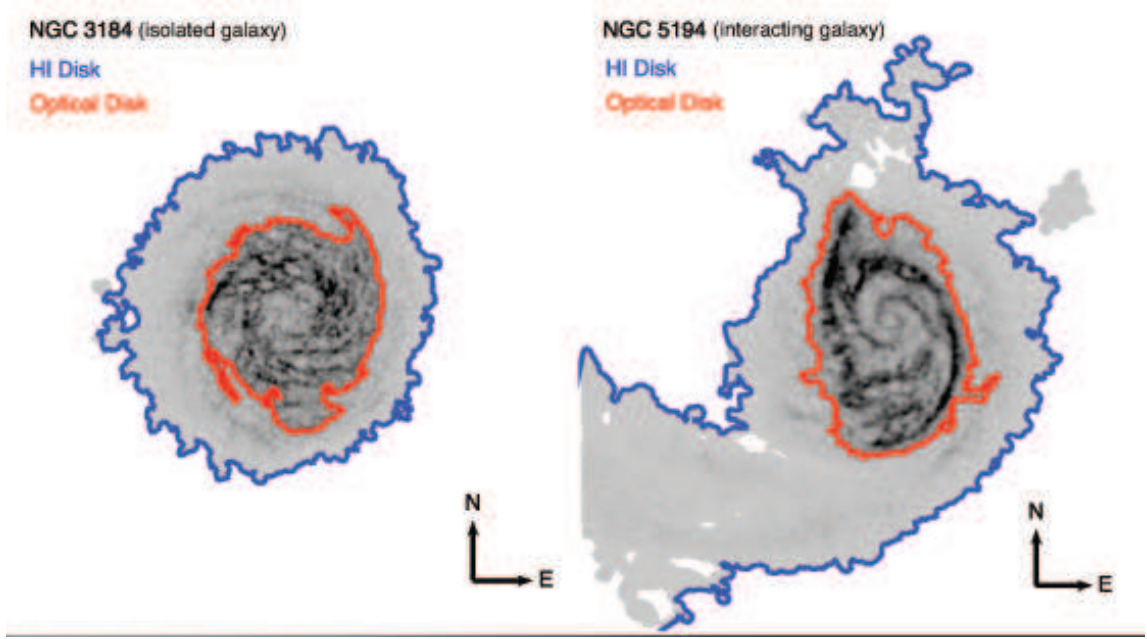}
\caption{\label{f:map}THINGS \hi \ maps (NA) for NGC 3184 and NGC 5194.  The edge of the optical or stellar disk is at $3 \times 10^{21}$ atoms cm$^{-2}$, indicated by the red contour, and the extent of the \hi \ disk ($10^{19}$ atoms cm$^{-2}$, the THINGS detection limit) is marked with the blue contour.}
\end{figure*}

\section{A case of two galaxies: NGC 3184 and NGC 5194}

\begin{figure*}
\centering
\includegraphics[angle=90, width=0.6\textwidth]{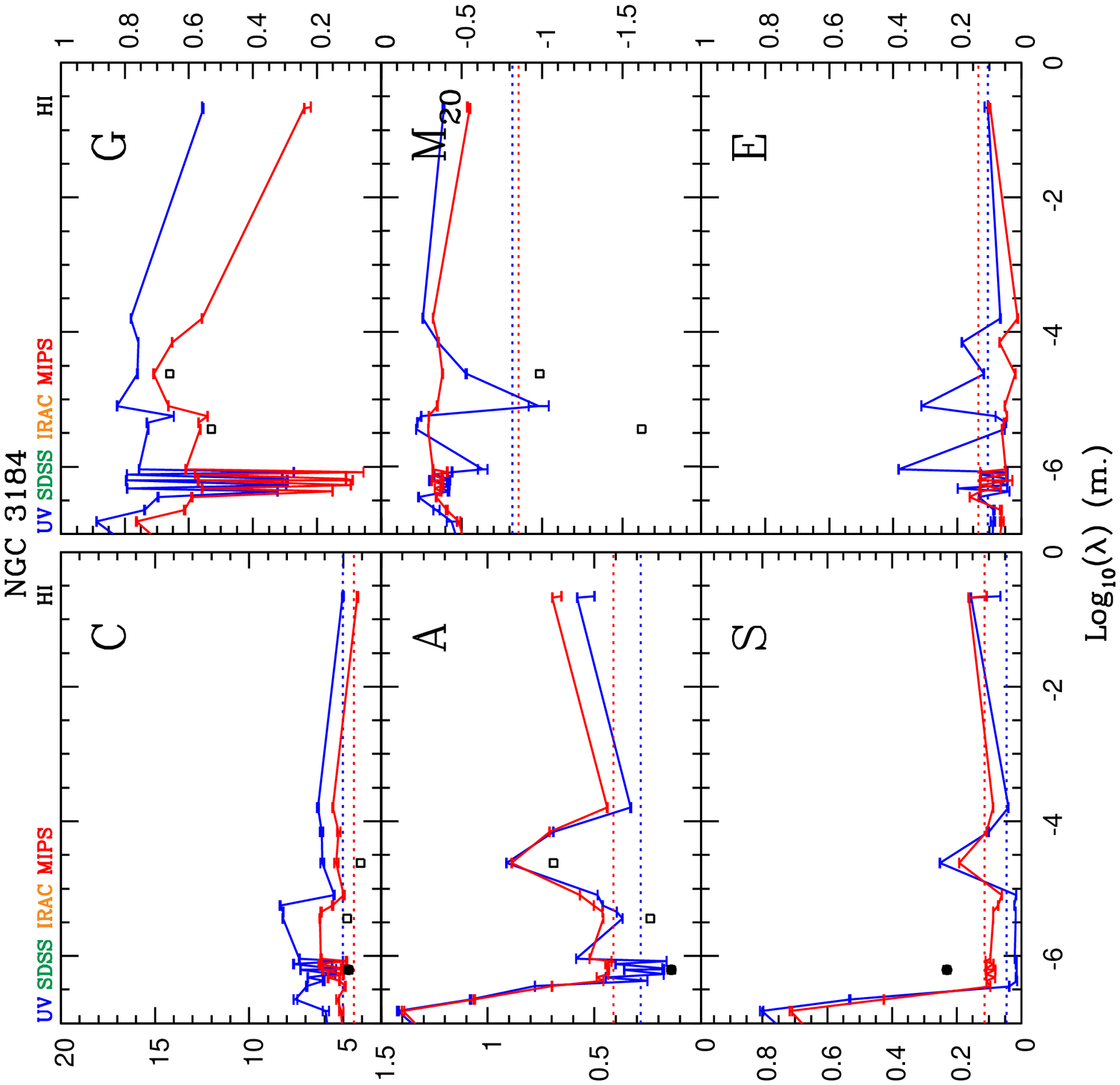}
\includegraphics[angle=90, width=0.6\textwidth]{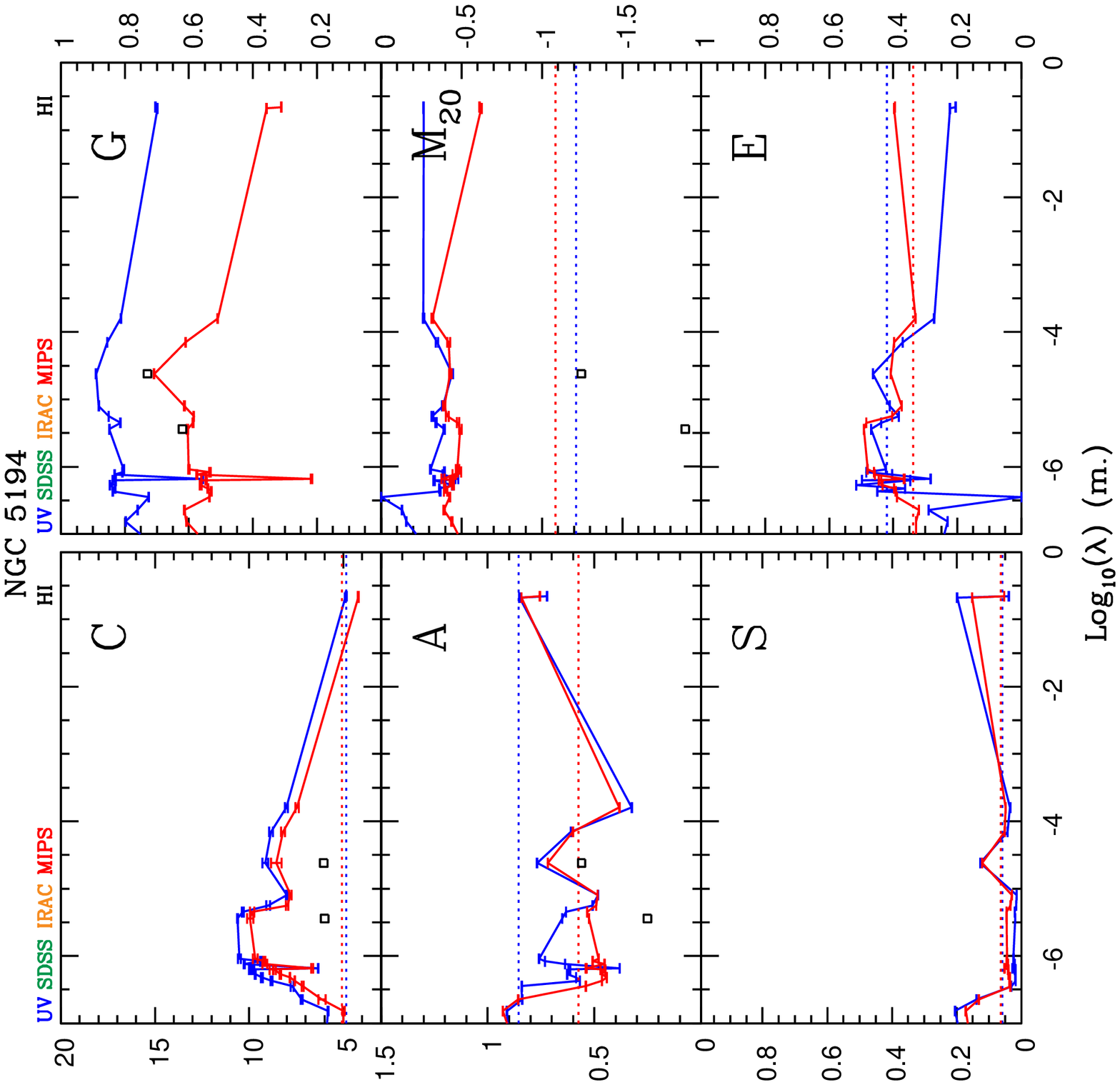}
\caption{\label{f:morph}The six structural parameters for the isolated NGC 3184 and interacting NGC 5194 as a function of wavelength.  The values for the optical disk (red) and those for the \hi \ disk (blue). The dashed lines are for the contours with equal pixel weights (set to 1) to serve as a reference (for Gini this value is 0, prefect equality). The optical values from \protect\cite{CAS} for NGC 3184 are indicated as black dots and the values for the 3.6 and 24 $\mu$m. \ from \protect\cite{Bendo07} are indicated with open squares. }
\end{figure*}

The above parameters were computed within two contours, one corresponding to approximately the extent of the stellar disk ($3 \times 10^{21}$ atoms/cm$^2$) and one to the extent of the \hi \ disk that could be observed with THINGS \citep{Walter08}, (Figure \ref{f:map}).
 We applied these contours on images of two galaxies, NGC 3184 and NGC 5194, spanning UV, optical, near and far-infrared from GALEX, SDSS and Spitzer \citep[see also][in preparation]{Holwerda10a}. The resulting morphology parameter values are plotted in Figure \ref{f:morph}, as computed in the stellar disk (red) and over the extent of the \hi \ disk (blue). 

We note that the area over which these parameters are computed does not matter much for many of the parameters but there is extra information at each wavelength. The exception is the Gini parameter, which increases significantly from the stellar disk to the \hi \ disk at all wavelengths. 
The parameters dependence on wavelength is more significant; Asymmetry, Smoothness and Gini are high in star-formation and 21 cm line emission and Concentration peaks in the optical.

\section{HI morphology as a interaction tracer}

In Figure \ref{f:morph} we can also compare the isolated, unperturbed galaxy NGC 3184 to the interacting NGC 5194 (M51). In the morphological parameters we can see that Concentration increases in the optical and near-infrared from isolated to interacting. Asymmetry decreases a little from isolated to interaction in UV and 24 $\mu$m emission (the star-formation tracers) but increases for \hi \ 21 cm emission. Both Gini and $M_{20}$ increase, notably in \hi. From Figure \ref{f:morph} we can conclude that there is as much signal of interaction in \hi \ as in the star-formation tracers (24 $\mu$m and UV) but any \hi \ classification scheme will have to be fine-tuned for this signal. Fortunately, we can use existing uniform \hi \ surveys such WHISP \citep{whisp} and THINGS \citep{Walter08} to test these morphology parameters \citep[see][all in preparation]{Holwerda10b,Holwerda10c,Holwerda10d}

\section{Use in Future Surveys}

The coming \hi \ surveys with the SKA precursors will produce a great number of resolved \hi \ datacubes, using a data-pipeline, producing data-cubes, moment order maps and a catalogue with dynamical ($W_{20}$, $W_{50}$, $V_{sys}$, $V_{rot}$ etc.) and flux information. Morphological parameters such as the above, CAS, Gini and $M_{20}$, of the zero-order moment map (the \hi \ surface density map) could be added to the catalogue, with the intention to expand the catalogue's parameter space such that first cuts can easily be made for isolated, harassed, and interacting galaxies.


\section{Applicability to data-cubes}


The application of morphological parameters to \hi \ is thusfar limited to the zero-order map, predominantly for comparison to other wavelengths but one strength of \hi \ data is its third dimension, frequency. Like the spatial dimensions of the data, this it has a finite sampling and range. In the discussions at the Panoramic Radio Astronomy conference, P. Serra mentioned the possibility of redefining these parameters for use on data-cubes. Following his suggestion, three of the above parameters could, in principle, easily be redefined to 3D morphology parameters: Asymmetry, Gini and $M_{20}$. 
In the case of asymmetry, the datacube would be mirrored around the central position ($x_c$, $y_c$, $f_c$) in space and frequency, the Gini parameter would simply have more input-pixels ranked by value (unfortunately most of those would be essentially noise), and the second order moment of a pixel i in the datacube would be redefined as: $M_i = I_i [(x_i - x_c)^2 + (y_i - y_c)^2 + (f_i - f_c)^2] $ with $M_{20}$ to follow \citep[see][]{Holwerda10d}. 
These three 3D morphology parameters, probably combined with zero-order map morphological parameters and dynamical indicators, could span a parameter space in which it could be possible to determine which datacube has unique \hi \ phenomena like warps, ``beards'' (anomalous velocity gas on one side of the disk) and of course interaction and harassment.


\begin{thebibliography}{15}
\expandafter\ifx\csname natexlab\endcsname\relax\def\natexlab#1{#1}\fi

\bibitem[{{Abraham} {et~al.}(1994){Abraham}, {Valdes}, {Yee}, \& {van den
  Bergh}}]{Abraham94}
{Abraham} R.~G., {Valdes} F., {Yee} H.~K.~C., {van den Bergh} S., 1994, \apj,
  432, 75

\bibitem[{{Abraham} {et~al.}(1996){Abraham}, {van den Bergh}, {Glazebrook},
  {Ellis}, {Santiago}, {Surma}, \& {Griffiths}}]{Abraham96a}
{Abraham} R.~G., {van den Bergh} S., {Glazebrook} K., {Ellis} R.~S., {Santiago}
  B.~X., {Surma} P., {Griffiths} R.~E., 1996, \apjs, 107, 1

\bibitem[{{Bendo} {et~al.}(2007){Bendo}, {Calzetti}, {Engelbracht},
  {Kennicutt}, {Meyer}, {Thornley}, {Walter}, {Dale}, {Li}, \&
  {Murphy}}]{Bendo07}
{Bendo} G.~J., {Calzetti} D., {Engelbracht} C.~W., {Kennicutt} R.~C., {Meyer}
  M.~J., {Thornley} M.~D., {Walter} F., {Dale} D.~A., {Li} A., {Murphy} E.~J.,
  2007, \mnras, 380, 1313

\bibitem[{{Conselice}(2003)}]{CAS}
{Conselice} C.~J., 2003, \apjs, 147, 1

\bibitem[{{Hibbard} {et~al.}(2001){Hibbard}, {van Gorkom}, {Rupen}, \&
  {Schiminovich}}]{Hibbard01}
{Hibbard} J.~E., {van Gorkom} J.~H., {Rupen} M.~P., {Schiminovich} D., 2001, in
  Astronomical Society of the Pacific Conference Series, Vol. 240, Gas and
  Galaxy Evolution, {Hibbard} J.~E., {Rupen} M., {van Gorkom} J.~H., eds., pp.
  657--+

\bibitem[{{Holwerda} {et~al.}(2010{\natexlab{a}}){Holwerda}, {Pirzkal}, {de
  Blok}, {Blyth}, {Bouchard}, {van der Heyden}, \& {Elson}}]{Holwerda10a}
{Holwerda} B.~W., {Pirzkal} N., {de Blok} W.~J.~G., {Blyth} S.-L., {Bouchard}
  A., {van der Heyden} K.~J., {Elson} E.~C., 2010{\natexlab{a}}, Quantified HI 
  Morphology I: Multiwavelength Morphology of NGC 3184 and NGC 5194, {\it in preparation}

\bibitem[{{Holwerda} {et~al.}(2010{\natexlab{b}}){Holwerda}, {Pirzkal}, {de
  Blok}, {Blyth}, {Bouchard}, {van der Heyden}, \& {Elson}}]{Holwerda10b}
---, 2010{\natexlab{b}}, Quantified HI Morphology II: Multiwavelength Morphology 
of the THINGS Galaxies, {\it in preparation}

\bibitem[{{Holwerda} {et~al.}(2010{\natexlab{c}}){Holwerda}, {Pirzkal}, {de
  Blok}, {Blyth}, {Bouchard}, {van der Heyden}, \& {Elson}}]{Holwerda10c}
---, 2010{\natexlab{c}}, Quantified HI Morphology III: Interaction, Dynamics 
and Star-Formation in the THINGS Galaxies, {\it in preparation}

\bibitem[{{Holwerda} {et~al.}(2010{\natexlab{d}}){Holwerda}, {Pirzkal}, {de
  Blok}, {Blyth}, {Bouchard}, {van der Heyden}, \& {Elson}}]{Holwerda10d}
---, 2010{\natexlab{c}}, Quantified HI Morphology IV: WHISP maps and cubes, {\it in preparation}

\bibitem[{{Kazantzidis} {et~al.}(2008){Kazantzidis}, {Bullock}, {Zentner},
  {Kravtsov}, \& {Moustakas}}]{Kazantzidis08}
{Kazantzidis} S., {Bullock} J.~S., {Zentner} A.~R., {Kravtsov} A.~V.,
  {Moustakas} L.~A., 2008, \apj, 688, 254

\bibitem[{{Lotz} {et~al.}(2004){Lotz}, {Primack}, \& {Madau}}]{Lotz04}
{Lotz} J.~M., {Primack} J., {Madau} P., 2004, \aj, 128, 163

\bibitem[{{Malin} \& {Hadley}(1997)}]{Malin97b}
{Malin} D., {Hadley} B., 1997, in Astronomical Society of the Pacific
  Conference Series, Vol. 116, The Nature of Elliptical Galaxies; 2nd Stromlo
  Symposium, {Arnaboldi} M., {Da Costa} G.~S., {Saha} P., eds., pp. 460--+

\bibitem[{{Scarlata} {et~al.}(2007){Scarlata}, {Carollo}, {Lilly}, {Sargent},
  {Feldmann}, {Kampczyk}, {Porciani}, {Koekemoer}, {Scoville}, {Kneib},
  {Leauthaud}, {Massey}, {Rhodes}, {Tasca}, {Capak}, {Maier}, {McCracken},
  {Mobasher}, {Renzini}, {Taniguchi}, {Thompson}, {Sheth}, {Ajiki}, {Aussel},
  {Murayama}, {Sanders}, {Sasaki}, {Shioya}, \& {Takahashi}}]{Scarlata07}
{Scarlata} C., {Carollo} C.~M., {Lilly} S., {Sargent} M.~T., {Feldmann} R.,
  {Kampczyk} P., {Porciani} C., {Koekemoer} A., {Scoville} N., {Kneib} J.-P.,
  {Leauthaud} A., {Massey} R., {Rhodes} J., {Tasca} L., {Capak} P., {Maier} C.,
  {McCracken} H.~J., {Mobasher} B., {Renzini} A., {Taniguchi} Y., {Thompson}
  D., {Sheth} K., {Ajiki} M., {Aussel} H., {Murayama} T., {Sanders} D.~B.,
  {Sasaki} S., {Shioya} Y., {Takahashi} M., 2007, \apjs, 172, 406

\bibitem[{{Takamiya}(1999)}]{Takamiya99}
{Takamiya} M., 1999, \apjs, 122, 109

\bibitem[{{van der Hulst} {et~al.}(2001){van der Hulst}, {van Albada}, \&
  {Sancisi}}]{whisp}
{van der Hulst} J.~M., {van Albada} T.~S., {Sancisi} R., 2001, in Astronomical
  Society of the Pacific Conference Series, Vol. 240, Gas and Galaxy Evolution,
  {Hibbard} J.~E., {Rupen} M., {van Gorkom} J.~H., eds., pp. 451--+

\bibitem[{{Walter} {et~al.}(2008){Walter}, {Brinks}, {de Blok}, {Bigiel},
  {Kennicutt}, {Thornley}, \& {Leroy}}]{Walter08}
{Walter} F., {Brinks} E., {de Blok} W.~J.~G., {Bigiel} F., {Kennicutt} R.~C.,
  {Thornley} M.~D., {Leroy} A., 2008, \aj, 136, 2563

\end{thebibliography}
\end{document}